\documentclass[12pt]{article}

\usepackage{graphicx}
\usepackage{amssymb}
\usepackage{amsmath}

\numberwithin{equation}{section}

\begin{document}
\baselineskip=16pt
\begin{titlepage}

\flushright{
MISC-2010-13}

\begin{center}
\vspace*{12mm}

{\Large\bf%
Negative mode of Schwarzschild black hole 
\\
from the thermodynamic instability
}\vspace*{10mm}

Takayuki Hirayama\footnote{e-mail: 
{\tt hirayama@cc.kyoto-su.ac.jp}}
\vspace*{4mm}

{\it Maskawa Institute for Science and Culture, 
Kyoto Sangyo University, 
\\
Kyoto, 603-8555, Japan
}\\[1mm]

\end{center}
\vspace*{10mm}

\begin{abstract}\noindent%
The thermodynamic instability, for example the negative heat capacity, of a black hole implies 
the existence of off-shell negative mode(s) (tachyonic mode(s)) around the black hole 
geometry in the Euclidean path integral formalism of quantum gravity.
We explicitly construct an off-shell negative mode inspired from the negative heat capacity 
in the case of Schwarzschild black hole with/without a cosmological constant. We carefully 
check the boundary conditions, i.e. the regularity at the horizon,  the traceless condition,
and the normalizability.
\end{abstract}

\end{titlepage}

\newpage


\section{Introduction}

The thermodynamics of black hole is a cornerstone in the search for a quantum gravity. The properties of black hole are expected to be captured by its thermodynamics, i.e. the temperature, mass, entropy, and other conserved charges. We then expect that, for example, the stability against small perturbations around the black hole will be equivalent with the stability as a thermodynamic system. In fact, it is known that a classical instability of black string which is called the Gregory-Laflamme instability~\cite{Gregory:1993vy} and the thermodynamic instability are equal in various black branes in Einstein general relativity, including the black holes discussed by Gregory and Laflamme~\cite{Gregory:1993vy}, black p-brane solutions in string theory~\cite{Hirayama:2002hn}, black strings in Anti-de Sitter 
space~\cite{Gubser:2000ec, {Hirayama:2001bi}}, D0-D2 bound state~\cite{Gubser:2004dr} and non-extremal smeared black branes~\cite{Harmark:2005jk}. Gubser and Mitra~\cite{Gubser:2000ec} conjectured they are equivalent when a black string has a non compact translational symmetry. But counter examples~\cite{Friess:2005zp} are also known where a scalar field, which does not possesses a conserved charge, expresses an instability which is not captured by the thermodynamics.
Another known fact is that an one-loop quantum instability of black hole, i.e. existence of non conformal negative mode(s)\footnote{
The conformal perturbations of the metric, which always decrease the Euclidean action and seem to render the path integral divergent, are decoupled and give no contribution to the path integral~\cite{Gibbons:1978ac}.
}, and the thermodynamic instability are also equal in various black holes in the path integral Einstein gravity~\cite{Gibbons:1976ue}. This is also checked for the black hole Anti de-Sitter space~\cite{Prestidge:1999uq} and rotating black holes~\cite{Monteiro:2009ke}, but a counter example~\cite{Hirayama:2008pf} is known in Einstein-Gauss-Bonnet theory.
These two known facts are related since the threshold mode of Gregory-Laflamme instability mode of a black string is equivalent with the non conformal negative mode~\cite{Gross:1982cv} of a black hole which appears as a slice of the black string. 

Reall~\cite{Reall:2001ag} gave an argument and it is now accepted that the thermodynamic instability implies the existence of non conformal negative mode(s). The existence of non conformal negative mode(s) is interpreted as the instability as the spontaneous nucleation of black holes in a hot flat space~\cite{Gross:1982cv, York:1986it} and thus it is important to understand a quantum gravity.
Reall gave an argument how a family of off-shell geometries around a black hole geometry is constructed, and discussed the existence of negative mode when the heat capacity is negative.
However as we will see soon, the off-shell modes constructed by Reall have problems. 
(i) The perturbation by taking the difference between two different off-shell geometries looks non-regular near the horizon in the Schwarzschild type coordinate system.
(ii) A finite cavity ($r=r_b$) is assumed and it is not clear whether $r_b\rightarrow\infty$ can be taken safely.
(iii) The traceless condition, i.e. the negative mode is a non conformal mode, cannot be satisfied at the horizon and at the boundary ($r=r_b$). Therefore we still have a question whether the negative heat capacity really implies the existence of non conformal negative mode. Here in this paper, we find a radial coordinate where the perturbations can be seen to be regular, improve the construction by Reall to satisfy (ii) and (iii), and explicitly construct a family of off-shell geometries around a black hole solution. We then show the existence of negative mode when the black hole shows the thermodynamic instability.

We briefly remind how Reall constructed the off-shell modes. The family of off-shell geometries
for a given temperature $T=1/\beta$ is parametrized by the horizon radius $r_h$. There the 
metric (Euclidean signature) is given
\begin{align}
 ds^2 &= g_{ab}dx^adx^b=U(r)dt^2+\frac{dr^2}{V(r)}+R^2(r)d\Omega_2^2
 .
\end{align}
When the geometry becomes a black hole geometry, the geometry extremizes the action. The boundary condition at the horizon $r=r_h$ ($U(r_h)=0$ and $V(r_h)=0$) which off-shell geometries should satisfy is $\sqrt{U'(r_h)V'(r_h)}=4\pi/\beta$, and that at the boundary $r=r_b$ is $\delta g_{tt}=0$.
Then put the black hole metric with the temperature $T'$ into the metric except for $U(r)$, and an arbitrary function but which satisfies the boundary conditions at the horizon and at the cavity into $U(r)$. Then this geometry is a off-shell mode parametrized by $r_h$ and we use the notation $U(r;r_h)$, $V(r;r_h)$ and $R(r;r_h)$. Notice that because of the boundary conditions for $U(r,r_h)$, the geometry is a off-shell (this geometry does not satisfy the equations of motion as long as $T'\neq T$.).

When we compute the perturbation by taking a difference between nearby two different off-shell geometries with the label $r_h$ and $r_h+\delta$, we naively obtain\footnote{$\partial_{r_h} F(r_h;r_h)$ means that we treat $F(r;r_h)$ as a function of $r$ and $r_h$ and take a derivative in terms of $r_h$.
}
\begin{align}
 \frac{1}{V(r;r_h+\delta)} -\frac{1}{V(r;r_h)}
 &=
 \frac{V'(r;r_h+\delta)^{-1}}{r-r_h-\delta}
 -  \frac{V'(r;r_h)^{-1}}{r-r_h}
 \nonumber\\ 
 &=
 \frac{V'(r;r_h)^{-1}}{r-r_h}\left( \frac{\delta}{r-r_h} - \delta \left(\partial_{r_h}V'(r;r_h)\right) V'(r;r_h)
 +{\cal O}(\delta^2)
 \right)
 ,
\end{align}
and then this asymptotic behaviour seems corresponding to a non regular mode because of $\delta/(r-r_h)$ 
behaviour as a perturbation. However this is an artifact due to the Schwarzschild coordinate system, and one can see the perturbations are regular at the horizon by using a different radial coordinate $y$ whose range is normalized to $y\in [0,1]$ using the construction 
in~\cite{Whiting:1988qr}. Therefore one should carefully choose a radial coordinate to explicitly construct off-shell modes suggested by Reall. 

One can see the traceless condition is not satisfied at the horizon and at the boundary since the traceless condition and the boundary conditions at the horizon
($\sqrt{U'(r_h)V'(r_h)}=4\pi/\beta$) and at the boundary ($\delta g_{tt}(r_b)=0$) are in general inconsistent. Also the normalizability near the boundary after taking the boundary infinity is not clear.
Thus in this paper, we discuss how to find a radial coordinate in which one can see the perturbations are regular at the horizon and are normalizable at the infinity after taking the boundary $r_b\rightarrow\infty$. Then we improve the construction of off-shell mode such that the perturbation satisfies the traceless condition. We then show the improved mode gives a negative mode. Another attempt to construct a non conformal negative mode by using a killing vector around the black hole geometry is given in the paper~\cite{Monteiro:2009tc}.


\section{Gravity action and black hole geometry}

In this section, we review the Einstein gravity action and the argument by Reall.
It is easy to understand our logic if we use an explicit example. Therefore we use a simple black hole solution, i.e. Schwarzschild black hole solution with/without a cosmological constant in four dimensions. It must be easy to generalize our argument to more general cases of black holes.

The Euclidean path integral of quantum gravity for the canonical ensemble and the physical Euclidean gravity action $I_p$ in a finite cavity are given
\begin{align}
 Z &= \int \!\! {\cal D}[g] \, e^{-I_P[g]}
 , \hspace{3ex}
 I_P[g] = I[g] - I_0 ,
 \\
 I[g] &= -\frac{1}{16\pi}\int_M \!\!\! d^4\! x \sqrt{g} \; \left( R-2\Lambda \right) 
 -\frac{1}{8\pi}\int_{\partial M} \!\!\! d^3\! x \sqrt{g_{(3)}} \; K ,
\end{align}
where $I_0$ is some reference action so that $I_P$ has a finite action,
$K$ is the Gibbons-Hawking surface terms,
$\partial M$ denotes the boundary.
The path integral is taken over Riemannian manifolds $(M,g)$ that are asymptotically flat or Anti-de Sitter space depending on the value of cosmological constant. The time direction should have a proper length $\beta=1/T$ where $T$ is the temperature. This path-integral is only well-defined in the semi-classical approximation, and the Einstein equations are
\begin{align}
 R_{ab} -\frac{1}{2} ( R-2\Lambda) g_{ab} &=0
 .
\end{align}
Around a solution of Einstein equations, the metric can be written
$g_{ab}=\bar{g}_{ab}+\delta g_{ab}$ and the action can be expanded around the 
solution as $I[g]=I_0[\bar{g}]+I_2[\bar{g},\delta g]$ where $I_2$ is quadratic in the 
fluctuation. The trace part of the metric perturbation has a wrong-sign kinetic term and is 
decoupled~\cite{Gibbons:1978ac}. The traceless part $h_{ab}$ gives
\begin{align}
 I_2 &= \int\! d^4x \sqrt{\bar{g}} h^{ab}\Delta_L h_{ab}
 ,
\end{align}
where $\Delta_L$ is called the Euclidean Lichnwerowicz operator. The perturbation should have a finite norm and regular everywhere including at the horizon. Then if the Lichnwerowicz operator has negative eigenvalue(s) $\lambda$,
\begin{align}
 \Delta_L h_{ab} &=\lambda h_{ab},
 \hspace{3ex}
 \lambda <0,
 \label{eigenvalue}
\end{align}
the solution is unstable. The mode with negative $\lambda$ is called a non-conformal negative mode.

On the other hand, we can demonstrate that a black hole with a thermodynamic instability have a negative mode. Since we are interested in a static black hole geometry, the metric ansatz is given
\begin{align}
 ds^2 &= U(r;r_h) dt^2 + \frac{1}{V(r;r_h)}dr^2 + R^2(r;r_h)d\Omega_2^2 ,
 \\
 \frac{4\pi}{\beta} &= \left.\sqrt{U'(r;r_h)V'(r;r_h)}\right|_{r=r_h} ,
\end{align}
where $r_h$ denotes the size of black hole horizon 
($U(r=r_h;r_h)=V(r=r_h;r_h)=0$)
and $U'(r,r_h)=\partial_r U(r,r_h)$ etc. (In the following, we often omit the index $r_h$ as long as it is clear.) With this ansatz the action becomes
\begin{align}
 I &\equiv I(r_h)
 \\
 &=
  - \frac{1}{16\pi}\int_M \! d^4x \sqrt{U(r)} \sqrt{g^{(3)}}( R^{(3)} - 2\Lambda ) 
 + \frac{1}{16\pi}\int_M \! d^4x \: 
 \partial_r \Big( \frac{\sqrt{V(r)}}{\sqrt{U(r)}}U'(r)R^2(r)|\cos\theta| \Big)
 \nonumber\\
 &\;\;\;\;
 - \frac{1}{16\pi}\int_{r=r_b} \!\!\!\! d^3x \: 2\sqrt{V(r)} \partial_r\Big(\sqrt{U(r)}R^2(r) \Big)|\cos\theta|
 \label{action}\\
 &=
 0
 - \frac{4\pi\beta}{16\pi} \frac{\sqrt{V(r)}}{\sqrt{U(r)}}U'(r)R^2(r)\Big|_{r=r_h}
 - \frac{4\pi\beta}{16\pi}4\sqrt{U(r)V(r)} R(r)R'(r)\Big|_{r=r_b}
  \\
 &= \beta H - S,
\end{align}
where $r_b$ is the boundary (the position of finite cavity), the entropy $S$ is the contribution from $r=r_h$ and $\beta H$, $H$ is Hamiltonian, is the rest of it in~\eqref{action}. $R^{(3)}$ is Ricci scalar constructed from the induced metric at a constant $t$.
The reference geometry is given by the same ansatz but no black hole, and we correspondingly have $\beta_0 H_0$ from $I_0$, i.e.
\begin{align}
 I_0 &=  \beta_0 H_0 =
 - \frac{4\pi\beta_0}{16\pi}4\sqrt{U(r;0)V(r;0)} R(r;0)R'(r;0)\Big|_{r=r'_b}
 ,
\end{align}
where $\beta_0$ and the boundary $r_b'$ are determined from the condition that two geometries have the same periodicity and the same radius of $S_2$, 
\begin{align}
 \beta_0 \sqrt{U(r'_b;0)} &= \beta \sqrt{U(r_b;r_h)} ,
 \\
 R(r'_b;0) &= R(r_b;r_h)
 .
\end{align}
We then denote $\beta H_P = \beta H - \beta_0 H_0$.

In this paper, we discuss Schwarzschild black hole in four dimensions with or without negative cosmological constant. When we use the Schwarzschild coordinate system, the black hole metric is given
\begin{align}
 U(r;r_h) = V(r;r_h) = f(r;r_h) \equiv 1-\frac{r_h}{r} 
 -\frac{\Lambda}{3} \Big( r^2 - \frac{r_h^3}{r} \Big) ,
 \hspace{3ex}
 R(r;r_h) = r ,
\end{align}
and then $R=4\Lambda$,
$\beta=4\pi r_h / ( 1 - \Lambda r_h^2) $,
\begin{align}
 \beta H &= \frac{1}{4} \beta ( -4r_b+\frac{4}{3}\Lambda r_b^3 +4r_h-\frac{4}{3}\Lambda r_h^3 )
 ,
 \\
 S &= \frac{1}{4}\beta ( r_h - \Lambda r_h^3 ) = \pi r_h^2
 ,
 \\
 \beta H_0 &= \frac{1}{4}\beta_0 ( -4r_b +\frac{4}{3}\Lambda r_b^3 )
 \\
 &=
 \frac{1}{4}\beta ( -4r_b + \frac{4}{3}\Lambda r_b^3 +2r_h -\frac{2}{3}\Lambda r_h^3 )
 +{\cal O}(r_b^{-1}),
 \\
 \rightarrow \beta H_P &= \beta \frac{r_h}{2} ( 1 - \frac{1}{3}\Lambda r_h^2)
 +{\cal O}(r_b^{-1}).
\end{align}
Then the Hawking temperature $T$, mass $M$ and entropy $S$ ($r_b\rightarrow\infty $) are
\begin{align}
 T &=\beta^{-1} = \frac{f'(r_h)}{4\pi} = \frac{1}{4\pi}\left(\frac{1}{r_h} -\Lambda r_h \right)
 , \hspace{3ex}
 M = \frac{r_h}{2} (1 -\frac{\Lambda}{3}r_h^2)
 , \hspace{3ex}
 S = \frac{1}{4} 4\pi r_h^2 .
\end{align}
Then the heat capacity
\begin{align}
 C_V &= \frac{dM}{dT} = -2\pi r_h^2\frac{ 1-\Lambda r_h^2}{1+\Lambda r_h^2}
\end{align}
is negative (positive) for $r_h < \sqrt{-\Lambda}$ ($r_h > \sqrt{-\Lambda}$). When the heat capacity is negative, the black hole is unstable as a thermodynamic system and we expect this thermodynamic instability appears as a negative mode in the semi-classical path integral of quantum gravity. In order to demonstrate that, one has to construct a family of geometries around the black hole solution and check the conditions, i.e. finite norm and regularity, are satisfied.

Reall discussed a series of geometries in the following way. Choose $V(r)$ and $R(r)$ as the black hole metric with the horizon $r=r_h+\delta$, and $U(r)$ is arbitrary except that $U(r)$ at the horizon $r=r_h+\delta$ and the boundary $r=r_b''$ (which is not necessary to be $r_b$) are chosen such that
\begin{align}
 \left. \sqrt{U'(r;r_h+\delta)V'(r;r_h+\delta)} \right|_{r=r_h+\delta}
 &= \frac{4\pi}{\beta} 
 ,
 \label{horizoncondition1}
 \\
 U(r_b'',r_h+\delta) = U(r_b'',r_h),
 \hspace{3ex}
 &R(r_b'',r_h+\delta) = R(r_b'',r_h) . 
 \label{horizoncondition2}
\end{align}
The first condition is necessary in order that the geometry avoids a conical singularity at the horizon $r=r_h+\delta$, and the second condition is that the boundary geometry should be kept fixed. Since $U(r)$ is not a black hole solution when $\delta\neq 0$, these geometries are off-shell geometries. In our case, we can write the series of geometries
\begin{align}
 U(r;r_h+\delta) &= f(r;r_h+\delta h(r) )
 ,\hspace{3ex}
 V(r;r_h+\delta) = f(r;r_h+\delta)
 ,\hspace{3ex}
 R(r;r_h+\delta) = r,
\end{align}
and $h(r)$ is an arbitrary function but satisfies the boundary conditions (\ref{horizoncondition1}) and (\ref{horizoncondition2}).
Then in our case, we have $r_b''=r_b$ and we obtain
\begin{align}
 I &\equiv I(r_h+\delta)
 \nonumber\\
 &=
 - \frac{1}{16\pi}\int_{r=r_h+\delta} \!\!\!\!\!\!\!\!\!\! d^3x \: \frac{\sqrt{V(r)}}{\sqrt{U(r)}}U'(r)R^2(r)|\cos\theta|
 - \frac{1}{16\pi}\int_{r=r_b} \!\!\!\!\!\!\! d^3x \: 4\sqrt{U(r)V(r)} R(r)R'(r)|\cos\theta|
 \nonumber\\
 &= -\frac{1}{16\pi} 4\pi\beta \frac{4\pi}{\beta}(r_h+\delta)^2
 +\frac{\beta}{4}\Big(
 -4r_b +\frac{4}{3}\Lambda r_b^3 + 4(r_h+\frac{\delta}{2})
 -\frac{4\Lambda}{3}( r_h^3 +\frac{3}{2}r_h^2\delta+\frac{3}{2}r_h\delta^2 +\frac{1}{2}\delta^3 )
 \Big)
 \nonumber \\
 &\;\;\;\;
 +{\cal O}(r_b^{-1})
\end{align}
and $I_0$ is same. We notice that the constraint equations are satisfied and then the action does not depend on $h(r)$.
Then 
\begin{align}
 I_P &= \beta H_P -S =\beta \frac{r_h+\delta}{2}(1-\frac{\Lambda}{3}(r_h+\delta)^2)
 -\frac{1}{4}4\pi(r_h+\delta)^2
 +{\cal O}(r_b^{-1})
 ,
\end{align}
and
\begin{align}
 I(r_h+\delta) - I(r_h) &= -\frac{\pi(1+\Lambda r_h^2)}{1-\Lambda r_h^2}\delta^2 +{\cal O}(\delta^3,r_b^{-1}) .
 \label{action2}
\end{align}
Therefore there is a tachyonic direction around the black hole solution when the heat capacity is negative. Since $U(r)$ is arbitrary between the horizon and the boundary, the trace can be zero by turning $U(r)$ and thus the negative mode is traceless except at the horizon and boundary.

In order that this off-shell modes really generate the non conformal negative mode, we have to check that the perturbations by taking the difference between the different off-shell modes satisfy the conditions, i.e. tracelessness, finite norm and regularity. We write the perturbations as follows,
\begin{align}
 ds^2 &= U(r)(1+\delta H_t(r)) dt^2 + \frac{1+\delta H_r(r)}{V(r)}dr^2 
 + R^2(r)(1+\delta H_\theta(r)) d\Omega_2^2,
\end{align}
and
\begin{alignat}{2}
 \delta H_t(r) &= \frac{U(r;r_h+\delta)}{U(r;r_h)} -1 &=&
 \frac{f(r;r_h+\delta h(r))}{f(r;r_h)} -1,
 \\
 \delta H_r(r) &= \frac{V(r;r_h)}{V(r;r_h+\delta)} -1 &=&
 \frac{f(r;r_h)}{f(r;r_h+\delta)} - 1,
 \\
 \delta H_\theta(r) &= \frac{R^2(r;r_h+\delta)}{R^2(r;r_h)} -1 &=&
 0 .
\end{alignat} 
Then $H_r(r)$ is divergent at $r=r_h+\delta$ before $r$ reaches the original horizon $r=r_h$ for $\delta>0$
and thus this perturbation seems not regular. Also since $\delta H_t(r)$, i.e. $h(r)$, should satisfy the boundary conditions (\ref{horizoncondition1}) and (\ref{horizoncondition2}) which are not consistent with the traceless condition in general, the traceless condition is not satisfied at the horizon and the boundary. For example at the boundary, we have 
$\delta H_t(r_b)=\delta H_\theta(r_b)=0$ instead $\delta H_r(r_b)\neq 0$ and thus the traceless condition is not satisfied.
Therefore we should find a different radial coordinate where one can check the perturbations are regular at the horizon and improve the construction to satisfy the traceless condition including at the horizon and the boundary.

We also should check whether we can safely take the limit $r_b\rightarrow\infty$. 
We thus check the normalizability which is given
\begin{align}
 \lim_{r_b\rightarrow\infty}
 \int \! d^4x \sqrt{g} \; \delta^2 (H_t(r)^2+H_r(r)^2+2H_\theta(r)^2)
 &<\infty
 .
 \label{norm}
\end{align}
In our case, the normalizability near the boundary gives
\begin{align}
 \int \! d^4x \sqrt{g} \; \delta^2 (H_t(r)^2+H_r(r)^2+2H_\theta(r)^2)
 &\sim \int^{r_b}_{r_h} dr r^2 \delta^2 \left[ \left(\frac{c}{r^3}\right)^2 +\left(\frac{c}{r^3}\right)^2 +0 \right]
 <\infty
\end{align}
where $c$ is a given number and then this is finite near the boundary.

\section{New coordinate}

Because of general covariance, it is not clear the perturbation is a regular or non-regular mode. To answer this question, it is better to define a better coordinate where everything becomes clear. The new coordinate $y$ should satisfy (i) the range is always fixed~\cite{Whiting:1988qr}
 (the original coordinate $r$ runs $r_h+\delta$ to $r_b$ which depends on the horizon $r_h+\delta$) and (ii) the normalizability near the boundary should be kept.

We find that it is not simple to find a new coordinate which satisfies (ii). For example, if we define $y=r-r_h$, this breaks (ii) in addition that the range $y\in [0,r_b-r_h-\delta]$ still depends on the horizon, 
since the metric and the normalizability become
\begin{align}
 &ds^2 = f(y+r_h;r_h) dt^2 +\frac{1}{f(y+r_h;r_h)} dy^2  +(y+r_h)^2 d\Omega_2,
 \\
 &\int \! d^4x \sqrt{g} \: \delta^2 H_y(y)^2
 \sim \int^{r_b-r_h-\delta} \!\!\!\!\!\!\!\!\! dy \: y^2 \delta^2 \left(\frac{1}{y}\right)^2 \sim r_b
 ,
\end{align}
for large $y\sim r_b$. Therefore the condition (ii) is easily broken. Another example which may be often used is 
\begin{align}
 y &= \frac{r-r_h}{r_b-r_h} ,
 \hspace{3ex}
 \Big({\rm i.e.}
 \hspace{1ex}
 r\equiv r(y;r_h) = r_h +(r_b-r_h)y \Big)
 \label{new1}
\end{align}
and the range does not depend on the horizon size, $y\in [0,1]$. However this coordinate again breaks the condition (ii), since the metric and the normalizability near the boundary $y\sim 1$ are
\begin{align}
 &ds^2 = f(r(y;r_h+\delta);r_h+\delta h(y)) dt^2 +\frac{r'(y;r_h+\delta)^2}{f(r(y);r_h+\delta)} dy^2 
 +r(y;r_h+\delta)^2 d\Omega_2,
 \\
 &\int \! d^4x \sqrt{g} \: \delta^2 H_y(y)^2
 \sim \int^{1} \!\! dy \: y^2 r_b^3 \delta^2 \left(\frac{1}{r_b}\right)^2
 \sim r_b.
\end{align}
This is divergent when we take $r_b\rightarrow\infty$.
Then we should use a more complicated coordinate. One such example is
\begin{align}
 y &= \frac{r}{r_b}\left( 1 - \left(\frac{r_h}{r}\right)^m\left(\frac{r_b-r}{r_b-r_h}\right)^n\right) ,
\end{align}
where $m\geq 0$ and $n\geq 0$, and the range $y$ is $y\in[0,1]$.
If we take $m=1$ and $n=1$, $y$ becomes (\ref{new1}). If we take $m=2$ and $n=1$, we find both the conditions (i) and (ii) are satisfied, since the normalizability near the boundary $y\sim 1$ becomes
\begin{align}
 \int \!\! d^4x \sqrt{g} \, \delta^2 H_y(y)^2
 \sim \int^{1} \!\!dy \, 
 y^2 r_b^3 \delta^2 \left(\frac{1}{r_b^2 y^2}\right)^2
 < \infty.
\end{align}
We will give a detail calculation in the next section.

\section{Non conformal negative mode}

Now we use the new coordinate and discuss whether the perturbations satisfy the regularity at the horizon and other conditions. The new coordinate as we discussed in the previous section is
\begin{align}
 y&=\frac{r}{r_b}\left( 1 - \left(\frac{r_h}{r}\right)^2\frac{r_b-r}{r_b-r_h}\right) ,
 &
 y &\in [0,1] ,
 \label{transe}
\end{align}
and thus
\begin{align}
 r &\equiv r(y;r_h)
 \\
 &=\frac{1}{2(r_b-r_h)}\left[
 -r_h^2+r_b(r_b-r_h)y
 +\sqrt{ r_h^2(2r_b-r_h)^2 -2r_h^2r_b(r_b-r_h)y +r_b^2(r_b-r_h)^2y^2 
 } 
 \right].
 \nonumber
\end{align}
The off-shell geometry with the label $r_h+\delta$ becomes
\begin{align}
 ds^2 &= U(y;r_h+\delta h(y)) dt^2 + \frac{1}{V(y;r_h+\delta)}dy^2 + R^2(y;r_h+\delta) d\Omega_2^2
 \\
 &= U(y;r_h)(1+\delta H_t(y)) dt^2 + \frac{1+\delta H_y(y)}{V(y;r_h)} dy^2 + R^2(y;r_h)(1+\delta H_\theta(y)) d\Omega_2^2,
 \label{metricpert}
 \\
 U(y;r_h) &= f(r(y;r_h);r_h) 
 , \hspace{3ex}
 V(y;r_h) = \frac{f(r(y;r_h);r_h)}{r'(y;r_h)^2}
 , \hspace{3ex}
 R(y;r_h) = r(y;r_h)
 .
\end{align}
Then the perturbations at the order $\delta$ near the horizon $y=0$ are given
\begin{alignat}{2}
 \delta H_t(y) &=
 -
 \frac{(2r_b^2+r_h^2)(1+\Lambda r_h^2)-2r_br_h(1+2\Lambda r_h^2)}{r_h(2r_b-r_h)(r_b-r_h)(1-\Lambda r_h^2)}
 h(0)
 +{\cal O}(y),
 \label{h1}
 \\
 \delta H_y(y) &=
 \frac{(2r_b^2+r_h^2)(1+\Lambda r_h^2)-2r_br_h(2+\Lambda r_h^2)}{r_h(2r_b-r_h)(r_b-r_h)(1-\Lambda r_h^2)}
  +{\cal O}(y),
 \label{h2} 
 \\
 \delta H_\theta(y) &=  
 \frac{2}{r_h}\delta+{\cal O}(y) ,
 \label{h3}
\end{alignat}
and then the boundary condition at the horizon (\ref{horizoncondition1}), which becomes $\delta H_t(y)=\delta H_y(y)$ at $y=0$, gives
\begin{align}
 h(0) = - \frac{(2r_b^2+r_h^2)(1+\Lambda r_h^2)-2r_br_h(2+\Lambda r_h^2)}{(2r_b^2+r_h^2)(1+\Lambda r_h^2)-2r_br_h(1+2\Lambda r_h^2)}
 =-1+{\cal O}(r_b^{-1})
 .
\end{align}

Thus this mode is a regular mode as seen below. If we solve the equation for the eigenfunction with negative $\lambda$ in (\ref{eigenvalue}) around the black hole solution, we obtain the asymptotic behaviours near the horizon
\begin{align}
 H_t(y) &= a_0 \left( \frac{1}{y} + a_1 \ln y +\cdots \right)
 +b_0 \left( 1 + b_1 y +\cdots \right),
 \\
 H_y(y) &= a_0 \left( -\frac{1}{y} + a_1 \ln y +\cdots \right)
 +b_0 \left( 1 +c_1 y +\cdots \right),
 \\
 H_\theta(y) &= -\frac{1}{2}\left( H_t(y) + H_r(y) \right),
\end{align}
where $a_0$ and $b_0$ are integration constants and other coefficients are determined from the equations of motion. Here the mode with $a_0\neq 0$ is a non regular mode, and the mode with $a_0=0$ is a regular mode and then the mode constructed above in (\ref{h1}), (\ref{h2}) and (\ref{h3}) is a regular mode.

\vspace{3ex}

As one can see from the asymptotic behaviour (\ref{h1}), (\ref{h2}) and (\ref{h3}), this perturbation is not traceless at the horizon. Also it is easy to find that this is not traceless at the boundary as well. Thus we here improve the perturbation by adding a gauge transformation, $\delta g'_{ab}=\delta g_{ab}+\nabla_a \xi_b+\nabla_b \xi_a$ with $\xi_y=\delta \xi(y)$ and other $\xi_a=0$, and then we obtain the metric perturbation
\begin{alignat}{2}
 \delta H_t(y) &= \frac{U(y,r_h+\delta h(y))}{U(y,r_h)} -1
 &&+ \delta\frac{V(y;r_h)}{2U(y;r_h)}U'(y;r_h)\xi(y),
 \label{met1}\\
 \delta H_y(y) &= \frac{V(y,r_h)}{V(y,r_h+\delta)} -1
 &&+\delta \Big( V(y;r_h)\xi'(y) + \frac{V'(y;r_h)}{2}\xi(y) \Big),
 \label{met2}\\
 \delta H_\theta(y) &=  \frac{R(y;r_h+\delta)^2}{R(y;r_h)^2}-1
 &&+ \delta\frac{V(y;r_h)}{R(y;r_h)}R'(y;r_h)\xi(y),
 \label{met3}
\end{alignat}
and $\xi(y)$ is chosen such that the traceless condition is always satisfied. We notice that our gauge is same as that in~\cite{Reall:2001ag} and the traceless and transverse condition is not the gauge condition, but a result of equations of motion. The boundary condition (\ref{horizoncondition1}) does not give a condition for $\xi(y)$, but the regularity gives a condition that $\xi(0)$ is finite. The other boundary condition (\ref{horizoncondition2}) gives a condition that $h(1)=\xi(1)=0$ and $\xi'(1)\neq 0$ to ensure the trace
$H_t(1)+H_y(1)+2H_\theta(1)=0$. We will later check that these conditions can be realized. Before that, we will compute the action to see whether this mode really gives a negative mode. Therefore we compute $I[g]=I[\bar{g}]+I_2[\bar{g},\delta g]$, and thus
\begin{align}
 I(r_h+\delta) &= I(r_h) -\frac{1}{32\pi} \int_M d^4x \sqrt{g} \; \delta \mbox{(Ein eq)} \; \delta g^{ab}
 ,
\end{align}
where $\delta$(Ein eq) is the linearlized Einstein equation around the black hole metric. We simply substitute \eqref{h1},\eqref{h2} and \eqref{h3}, we obtain\footnote{
In the real computation, we again go back to the original coordinate $r$ and compute the action. We give the detail computation in the Appendix.}
\begin{align}
 I(r_h+\delta) &= I(r_h) 
 -\delta^2\frac{\beta}{8} \int_{0}^{1} \! dy
 \Big( A[h''(y),h'(y),h(y)]
 \nonumber\\
 &\;\;\;\;
 + B_1[h''(y),h'(y),h(y),\xi(y)] + B_2[h(y),h'(y),\xi'(y)]+ C[\xi(y)^2] \Big)
 +{\cal O}(\delta^3).
\end{align}
The terms $C$ are quadratic in terms of $\xi(y)$, the terms $B_1$ and $B_2$ are linear terms in terms of $\xi(y)$ or $\xi'(y)$ and the terms $A$ are the rest. $C[\xi(y)^2]=0$ since $\xi$ is a gauge freedom and the mixing terms $B_1$ and $B_2$ are zero except at the boundaries $y=0$ and $y=1$ since $\xi$ is again a gauge freedom (zero mode). 
In fact we can rewrite
\begin{align}
 B_1[h''(y),h'(y),h(y),\xi(y)] + B_2[h(y),h'(y),\xi'(y)] &= \partial_y B[h'(y),h(y),\xi(y)] ,
\end{align}
and we obtain $B[h'(0),h(0),\xi(0)]=0$ (because $y=0$) and $B[h'(1),h(1),\xi(1)]=0$ (because $\xi(1)=0$).
Now we compute $A[h''(y),h'(y),h(y)]$. In our case, we can integrate along $y$ direction (since the constraint equations at a constant $t$ are satisfied), and we obtain
\begin{align}
 -\frac{\delta^2\beta}{8} \int_0^1 \! dy \; A[h''(y),h'(y),h(y)]
 &=-\frac{\delta^2\pi (1+\Lambda r_h^2)}{1-\Lambda r_h^2}
 +{\cal O}(r_b^{-1})
 ,
 \label{aaaaa}
\end{align}
where $\beta=4\pi r_h/(1-\Lambda r_h^2)$ and the asymptotic forms of $h(y)$
\begin{align}
 h(y) &= - \frac{(2r_b^2+r_h^2)(1+\Lambda r_h^2)-2r_br_h(2+\Lambda r_h^2)}
 {(2r_b^2+r_h^2)(1+\Lambda r_h^2)-2r_br_h(1+2\Lambda r_h^2)} + {\cal O}(y) ,
 \\
 h(y) &= 0 + h_1 \left( \frac{1}{y} - 1 \right)  + h_2 \left( \frac{1}{y} - 1 \right)^2
 +{\cal O}((y^{-1}-1)^3) ,
\end{align}
are used. In summary we have
\begin{align}
 I(r_h+\delta) - I(r_h)
 &= -\frac{\delta^2\pi (1+\Lambda r_h^2)}{1-\Lambda r_h^2}
 +{\cal O}(r_b^{-1},\delta^3).
 \label{action3}
\end{align}
Thus as long as $r_h^2<1/(-\Lambda)$, this mode gives a negative mode!

\bigskip

Now we go back to determine $\xi(r)$ in order to see if the traceless condition can really be satisfied. The traceless condition gives a differential equation for $\xi(y)$, (explicit form is written in Appendix C),
\begin{align}
 \xi'(y) &= F[h(y),\xi(y)]
 .
\end{align}
We solve this differential equation near the horizon and the boundary and obtain
\begin{align}
 \xi(y) &= a_0 y^{-1} + a_1 + a_2y +{\cal O}(y^2),
 \label{asym2}\\
 \xi(y) &= b_0 +b_1(y^{-1}-1) +b_2(y^{-1}-1)^2 +{\cal O}((y^{-1}-1)^3) ,
 \label{asym3}
\end{align}
where we have used the asymptotic solution of $h(y)$.
The integration constants are $a_0$ and $b_0$. The boundary condition and the regularity impose both $a_0$ and $b_0$ are zero. However even we take $a_0=0$ at the horizon, we in general have nonzero $b_0$ after solving the differential equation from the horizon to the boundary for a given $h(y)$. However since $h(y)$ is arbitrary between the horizon and the boundary, we can use this freedom to realize both $a_0$ and $b_0$ are zero. Therefore we can realize the traceless condition and do not break the regularity and normalizability.

\bigskip

We finally check the normalizability near the boundary. The norm is given
\begin{align}
 \int \!\! d^4x \sqrt{g} \, ( H_t(y)^2+H_y(y)^2+2H_\theta(y)^2 ) =N.
\end{align}
Near $y=1$ we obtain
\begin{align}
 \int \!\! d^4x \sqrt{g} \, ( H_t(y)^2+H_y(y)^2+2H_\theta(y)^2 )
 \sim
 \int_0^1 \! \! dy \, 
 \frac{r_h^2(y-1)^2(y^4-4y^3+9y^2-14y+10)}{y^2 r_b} <\infty
\end{align}
and then this mode is normalizable. In order to compute the eigenvalue in (\ref{eigenvalue}), we have to compute the value $N$ and using (\ref{action3}) or (\ref{action2}),
\begin{align}
 \lambda = - \frac{1}{N} \frac{\pi(1+\Lambda r_h^2)}{1-\Lambda r_h^2}
 .
\end{align}
We here did not give the explicit form of $h(y)$ and cannot compute the value $N$.

In summary, we explicitly constructed a non-conformal negative mode inspired from the black hole thermodynamic instability. Our mode satisfies all the boundary conditions and normalizability.


\section{Summary}

In this paper, we explicitly constructed a negative mode around the black hole geometry when it has a thermodynamic instability. The negative mode satisfies the boundary conditions at the horizon and the boundary, i.e. regularity, traceless and normalizability. It is important to find a proper coordinate system and we believe it is easy to generalize our arguments to the case of more general black holes, such as rotating and charged black holes in four and higher dimensions.

We did not impose the transverse condition which is a result of Einstein equation in this gauge. (Notice that we use the gauge used in~\cite{Reall:2001ag} and the transverse condition is obtained as a equation of motion.) We also did not compute the eigenvalue. We have to fine tune the function $h(y)$ between the horizon and the boundary in order to have a normalizable mode and then did not minimize the norm. It is interesting to compute the eigenvalue to compare our negative mode with the negative mode obtained by solving the eigenvalue equation.

In many cases, the negative mode disappears exactly when the thermal instability disappears. There are however counter examples. Therefore it is interesting if we can prove when the negative mode implies the thermodynamic instability.

\subsection*{Acknowledgments}

The author would like to thank all the members of the string group in
Taiwan and all the members in Maskawa Institute. 
This work is supported by Maskawa Institute in Kyoto Sangyo University.

\appendix

\section{Action}

\begin{align}
 I &= -\frac{1}{16\pi}\int_M \!\!\! d^4\! x \sqrt{g} \; \left( R-2\Lambda \right) 
 -\frac{1}{8\pi}\int_{\partial M} \!\!\! d^3\! x \sqrt{g_{(3)}} \; K ,
\end{align}
Since we are interested in a static black hole geometry, the metric ansatz is given
\begin{align}
 ds^2 &= U(r)(1+\delta H_t(r)) dt^2 + \frac{1+\delta H_r(r)}{V(r)}dr^2 
 + R^2(r)(1+\delta H_\theta(r))d\Omega_2^2 ,
\end{align}
With this ansatz the action becomes
\begin{align}
 I 
 &=
  - \frac{1}{16\pi}\int_M \!\! \sqrt{U} \sqrt{g^{(3)}}( R^{(3)} - 2\Lambda ) 
 + \frac{1}{16\pi}\int_M \!\!
 \partial_r \Big( \frac{\sqrt{V}}{\sqrt{U}}U'R^2|\cos\theta| \Big) 
 \nonumber
 \\ 
  &\;\;\;
 +\frac{\delta}{16\pi}\int_M\!\! \partial_r\Big( \frac{\sqrt{V}R|\cos\theta|}{2\sqrt{U}}
 (RU' H_t +2RU H_t' -RU' H_r -4UR' H_r +4UR' H_\theta +4RU H_\theta' ) \Big)
 \nonumber
 \\
 & \;\;\;
 -\frac{1}{16\pi} \int_M\!\!\! \sqrt{g}\; 
 \Big(R_{ab}-\frac{1}{2}(R-2\Lambda)\Big)\delta g^{ab}
 +{\cal O}(\delta^2).
\end{align}
Then since at the horizon, we have $U=V=0$ and $H_t=H_r$, the boundary terms cancel out and then we do not need the boundary action.

\section{Detail of the computation}

In the computation of action, it is easier to go back to the original coordinate $r$ instead of using $y$ coordinate (\ref{transe}). We transform the metric perturbations (\ref{metricpert}) with (\ref{met1}), (\ref{met2}) and (\ref{met3}), and we obtain
\begin{align}
 ds^2 &= U(r;r_h)(1+\delta H_t(r))dt^2 + \frac{1+\delta H_y(r)}{V(r;r_h)} dr^2 
 +R^2(r;r_h)(1+\delta H_\theta(r)) d\Omega_2^2,
 \\
 U(r;r_h) &= V(r;r_h)=f(r;r_h), \hspace{3ex} R(r;r_h)=r,
\end{align}
and
\begin{align}
 H_t(r) &= \frac{A(r)}{B(r)} +\frac{V(r;r_h)}{2U(r;r_h)}U'(r;r_h)\xi(r) ,
 \\
 A(r) &=
 \Big(  \left( (3+\Lambda {r_h}^{2})(r+r_h)+4 r_h \Lambda {r}^{2} \right) {r_b}^{2}
 + \left( -2 \Lambda {r_h}^{4}-4 r_h \Lambda {r}^{3}
 -6 {r_h}^{2}\Lambda {r}^{2}-6 r_h r \right) r_b
 \nonumber\\
 &\,\,\,\,\,\,
 -\Lambda {r_h}^{4}r+2 \Lambda {r_h}^{3}{r}^{2}+3 {r_h}^{2}r
 +2 \Lambda {r}^{3}{r_h}^{2} \Big) h(r) ,
 \\
 B(r)&=
 \left( \Lambda {r_h}^{2}+\Lambda r_h r-3+\Lambda {r}^{2} \right)  
 \left( r_b-r_h \right)  \left(  \left( {r}^{2}+{r_h}^{2} \right) r_b-{r}^{2}r_h \right) ,
\end{align}
\begin{align}
 H_y(r) &= \frac{C(r)}{D(r)} +\Big( V(r;r_h)\xi'(r)+\frac{V'(r;r_h)}{2}\xi(r)
 \Big) ,
 \\
 C(y) &=
 \left( -8 \Lambda r^4 r_h-3 r_h^2 r
 -5\Lambda r^3 r_h^2
 -3r^3
 -5\Lambda r_h^3 r^2
 +9 r^2 r_h
 +3\Lambda r_h^5
 -15 r_h^3
 +3\Lambda r_h^4 r \right) r_b^3
 \nonumber\\
 &\,\,\,\,
 +
 \left( 30 r_h^3r
 -10\Lambda r_h^5r
 +16\Lambda r_h^2 r^4
 +4 r_h \Lambda {r}^{5}
 +5 \Lambda {r_h}^{4}{r}^{2}
 +9 {r}^{3}r_h
 +6 {r_h}^{4}
 -15 {r}^{2}{r_h}^{2}
 +3 \Lambda {r}^{3}{r_h}^{3} \right) {r_b}^{2}
 \nonumber
 \\
 &\,\,\,\,
 + 
 \left( -9 {r}^{3}{r_h}^{2}
 +6 {r}^{2}{r_h}^{3}
 -2 \Lambda {r_h}^{5}{r}^{2}
 -15 r {r_h}^{4}
 -10 \Lambda {r_h}^{3}{r}^{4}
 -6 {r_h}^{2}\Lambda {r}^{5}
 +5 \Lambda {r_h}^{6}r
 +\Lambda {r_h}^{4}{r}^{3} \right) r_b
 \nonumber\\
 &\,\,\,\,
 +
 2 \Lambda {r}^{5}{r_h}^{3}
 +3 {r}^{3}{r_h}^{3}
 -\Lambda {r_h}^{5}{r}^{3}
 +2 \Lambda {r_h}^{4}{r}^{4},
 \\
 D(r) &=\left( \Lambda {r_h}^{2}
 +\Lambda r_h r-3+\Lambda {r}^{2} \right)  \left( r_b-r_h \right)  
 \left(  \left( {r}^{2}+{r_h}^{2} \right) r_b-{r}^{2}r_h \right) ^{2},
 \\
 H_\theta(r) &=
 2 {\frac { \left( 2 {r_b^{2}+ \left( -2 r-r_h \right) r_b+r_h} r \right) r_h}{ \left( r_b-r_h
 \right)  \left(  \left( r^{2}+{r_h}^{2} \right) r_b-{r}^{2}r_h \right) }}
 +\frac{V(r;r_h)}{R(r;r_h)}R'(r;r_h)\xi(r)
 .
\end{align}
We can then relatively easily compute $A$ in (\ref{aaaaa}),
\begin{align}
 &-\frac{\beta}{8}\int_0^1\! \! dy \,  A[h''(r),h'(r),h(r)] =E(r) h'(r)+F(r)h(r)+G(r)\Big|_0^1 .
 \\
 &E(r)=\frac{E_1(r)}{E_2(r)},
 \\
 &E_1(r)=  r {r_h} \beta
 \left( r-r_h \right)  \left( r_b-r \right) 
 \left( 2 r_b-r_h \right)  \Big(  \left( \Lambda {r_h}^{3}
 +3 r_h+\Lambda {r_h}^{2}r+4 r_h {\Lambda} {r}^{2}+3 r \right) {r_b}^{2}
 \nonumber\\
 & \,\,\,\,\,\,\,\,\,\,\,\,\,\,\,\,\,\,\,
 - \left( 2 {\Lambda} {r_h}^{4} +4 r_h \Lambda {r}^{3}
 +6 {r_h}^{2}\Lambda {r}^{2} +6 r_h r \right) r_b
 +( 3 +\Lambda (2r^2+2r_h r -r_h^2) )r_h^2 r
 \Big),
 \\
 &E_2(r)=
 6\left(  \left( {r}^{2}+{r_h}^{2} \right) r_b-{r}^{2}r_h \right) ^{2} 
 \left( r_b-r_h \right) ^{2}.
\end{align}
\begin{align}
 &F(r) = \frac{F_1(r)}{F_2(r)},
 \\
 &F_1(r)= \Big( 
 47 {r}^{4}{\Lambda}^{2}{r_h}^{5}+108 \Lambda {r_h}^{5}{r}^{2}
 -30 {r_h}^{2}\Lambda {r}^{5}+3 {{\Lambda}}^{2}{r_h}^{9}
 +2 {r}^{2}{\Lambda}^{2}{r_h}^{7}+16 {r}^{6}{\Lambda}^{2}{r_h}^{3}
 +7 {r}^{5}{{\Lambda}}^{2}{r_h}^{4}
 \nonumber
 \\
 &-126 {r}^{2}{r_h}^{3}-48 {r}^{6}{\Lambda} r_h
 -60 \Lambda {r_h}^{4}{r}^{3}-9 {r_h}^{5}
 +6 \Lambda {r_h}^{3}{r}^{4}-9 {r}^{5}+6 \Lambda {r_h}^{7}
 +26 {r}^{3}{\Lambda}^{2}{r_h}^{6}+18 {\Lambda} {r_h}^{6}r
 \nonumber
 \\
 &
 +3 r {\Lambda}^{2}{r_h}^{8}+18 {r}^{3}{r_h}^{2}+27 r {r_h}^{4}+27 {r}^{4}r_h
 -32 {r}^{7}{r_h}^{2}{\Lambda}^{2} \Big) {\beta}
 r_b^5
 \nonumber\\
 &
 -r_h  \Big( 6 \Lambda {r_h}^{6}r-198 {r}^{2}{{r_h}}^{3}+63 r {r_h}^{4}-9 {r}^{5}-9 {r_h}^{5}
 +252 {\Lambda} {r_h}^{4}{r}^{3}-240 \Lambda {r_h}^{3}{r}^{4}
 +54 {r_h}^{2}\Lambda {r}^{5}
 \nonumber
 \\
 &+185 {r}^{4}{\Lambda}^{2}{r_h}^{5}+135 {r}^{4}r_h-198 {r}^{3}{r_h}^{2}+168 
 \Lambda {r_h}^{5}{r}^{2}-264 {r}^{6}\Lambda r_h
 +30 {r}^{2}{\Lambda}^{2}{r_h}^{7}-6 {r}^{3}{\Lambda}^{2}{r_h}^{6}
 \nonumber \\
 &
 -64 {r}^{7}{r_h}^{2}{\Lambda}^{2}-48 {r}^{7}\Lambda
 +51 {r}^{5}{\Lambda}^{2}{r_h}^{4}-48 {r}^{8}r_h {\Lambda}^{2}
 +11 r {\Lambda}^{2}{r_h}^{8}+120 {r}^{6}{\Lambda}^{2}{r_h}^{3}
 +9 {\Lambda}^{2}{{r_h}}^{9} \Big) {\beta}
 r_b^4
 \nonumber
 \\
 &+{r_h}^{2} \Big( -112 {r}^{8}r_h {\Lambda}^{2}+32 {r}^{7}{r_h}^{2}{\Lambda}^{2}
 +200 {r}^{6}{\Lambda}^{2}{{r_h}}^{3}+110 {r}^{5}{\Lambda}^{2}{r_h}^{4}
 +173 {r}^{4}{\Lambda}^{2}{r_h}^{5}+45 r {r_h}^{4}
 +36 {r}^{5}
 \nonumber\\
 &
 -36{r}^{2}{r_h}^{3}+6 {\Lambda}^{2}{r_h}^{9}-34 {r}^{3}{\Lambda}^{2}{r_h}^{6}
 +462 \Lambda {r_h}^{4}{r}^{3}+54 {r_h}^{2}\Lambda {r}^{5}
 -336 {r}^{6}\Lambda r_h-126 \Lambda {r_h}^{3}{r}^{4}
 +117 {r}^{4}{r_h}
 \nonumber\\
 &
 -432 {r}^{3}{r_h}^{2}-6 \Lambda {r_h}^{7}+30 {\Lambda} {r_h}^{5}{r}^{2}
 +50 {r}^{2}{\Lambda}^{2}{{r_h}}^{7}-216 {r}^{7}\Lambda
 +5 r {\Lambda}^{2}{r_h}^{8}-6 \Lambda {r_h}^{6}r-16 {r}^{9}{\Lambda}^{2}
 \Big) {\beta}
 r_b^3
 \nonumber\\
 &+
 {r_h}^{3}r  \Big( 45 {r}^{3}r_h+270 {r}^{2}{r_h}^{2}-9 {r_h}^{4}-54 {r_h}^{3}r-72 {r}^{4}
 +312 {r}^{6}{\Lambda}+108 {r}^{5}r_h \Lambda-128 {r}^{5}{\Lambda}^{2}{r_h}^{3}
 \nonumber\\
 &
 -96 {r}^{3}{r_h}^{3}\Lambda-61 {r}^{3}{{\Lambda}}^{2}{r_h}^{5}+42 r \Lambda {r_h}^{5}
 -132{r}^{6}{\Lambda}^{2}{r_h}^{2}+30 {r}^{2}{\Lambda}^{2}{r_h}^{6}
 -252 {r}^{2}\Lambda {r_h}^{4}-24 r {{\Lambda}}^{2}{r_h}^{7}
 \nonumber
 \\
 &
 +32 {\Lambda}^{2}{r}^{8}+92 {r}^{7}{r_h} {\Lambda}^{2}+{\Lambda}^{2}{r_h}^{8}
 -80 {r}^{4}{\Lambda}^{2}{r_h}^{4}-24 {r}^{4}{r_h}^{2}{\Lambda} \Big) {\beta}
 r_b^2
 \nonumber
\end{align} 
\begin{align} 
 &+
 {r_h}^{4}{r}^{2} \Big( 45 {r}^{3}+18 {r_h}^{3}-54 {{r_h}}^{2}r
 -72 {r}^{2}r_h+4 {r}^{2}{r_h}^{5}{\Lambda}^{2}+72 \Lambda {r_h}^{3}{r}^{2}
 +24 {r}^{4}r_h {\Lambda}+2 {\Lambda}^{2}{r_h}^{7}-12 \Lambda {{r_h}}^{5}
 \nonumber\\
 &
 +36 {r}^{4}{r_h}^{3}{\Lambda}^{2}+92 {r}^{5}{{r_h}}^{2}{\Lambda}^{2}
 +48 \Lambda {r_h}^{4}r-10 r {{r_h}}^{6}{\Lambda}^{2}-32 {r}^{6}{\Lambda}^{2}r_h
 -20 {r}^{7}{\Lambda}^{2}-180 \Lambda {r}^{5}+30 {\Lambda} {r}^{3}{r_h}^{2}
 \nonumber\\
 & 
 +9 {r}^{3}{r_h}^{4}{\Lambda}^{2} \Big) {\beta}
 r_b
 \nonumber\\
 &+{r}^{4}{r_h}^{5} \Big( 36 \Lambda {r}^{3}+4 {r}^{5}{{\Lambda}}^{2}
 +18 r_h-9 r+2 {\Lambda}^{2}{r_h}^{5}-12 r \Lambda {r_h}^{2}
 +4 {r}^{4}{\Lambda}^{2}{r_h}-12 \Lambda {r_h}^{3}+5 r {\Lambda}^{2}{r_h}^{4}
 \nonumber
 \\
 &
 -4 {r}^{2}{r_h}^{3}{\Lambda}^{2}-12 {r}^{2}r_h \Lambda-20 {r}^{3}{r_h}^{2}{\Lambda}^{2} 
 \Big) {\beta},
\nonumber\\
&F_2(r)=
12  \left( {r}^{2}{r_b}-{r}^{2}r_h+{r_h}^{2}{r_b}
 \right) ^{3} \left( {r_b}-r_h \right) ^{2} \left( {\Lambda} {r_h}^{2}+\Lambda r_h r-3
 +\Lambda {r}^{2}
 \right) .
\end{align}
\begin{align}
 &G(r)=
 - \beta\Big( 47 {r}^{4}{\Lambda}^{2}{r_h}^{5}+108 \Lambda {r_h}^{5}{r}^{2}
 -30  {r_h}^{2}\Lambda {r}^{5}+3 {{\Lambda}}^{2}{r_h}^{9}
 +2 {r}^{2}{\Lambda}^{2}{r_h}^{7}+16 {r}^{6}{\Lambda}^{2}{r_h}^{3}
 +7 {r}^{5}{{\Lambda}}^{2}{r_h}^{4}
 \nonumber\\
 &
 -126 {r}^{2}{r_h}^{3}-48 {r}^{6}{\Lambda} r_h-60 \Lambda {r_h}^{4}{r}^{3}
 -9 {r_h}^{5}+6 \Lambda {r_h}^{3}{r}^{4}-9 {r}^{5}+6 \Lambda {r_h}^{7}
 +26 {r}^{3}{\Lambda}^{2}{r_h}^{6}+18 {\Lambda} {r_h}^{6}r
 \nonumber\\
 &
 +3 r {\Lambda}^{2}{r_h}^{8}
 +18 {r}^{3}{r_h}^{2}+27 r {r_h}^{4}+27 {r}^{4}r_h
 -32 {r}^{7}{r_h}^{2}{\Lambda}^{2} \Big) 
 r_b^7
 \nonumber\\
 &
 -\beta r_h  \Big( 504 {r}^{6}\Lambda r_h-189{r}^{4}r_h+27 {r_h}^{5}+48 {r}^{7}\Lambda
 -117 r {r_h}^{4}+450 {r}^{2}{r_h}^{3}-42 \Lambda {r_h}^{6}r
 +48 \Lambda {r_h}^{5}{r}^{2}
 \nonumber\\
 &
 +4 r C {\Lambda} {r_h}^{7}+12 {r}^{5}C {r_h}^{3}{\Lambda}
 +4 {r}^{7}C \Lambda r_h+12 {r}^{3}C \Lambda {r_h}^{5}+16 {r}^{2}C {\Lambda} {r_h}^{6}
 +24 {r}^{4}C {r_h}^{4}\Lambda
 \nonumber\\
 &
 +16{r}^{6}C {r_h}^{2}\Lambda+27 {r}^{5}+660 {\Lambda} {r_h}^{3}{r}^{4}
 -567 {r}^{4}{\Lambda}^{2}{r_h}^{5}-226 {r}^{2}{\Lambda}^{2}{r_h}^{7}
 -190 {r}^{3}{{\Lambda}}^{2}{r_h}^{6}-65 r {\Lambda}^{2}{r_h}^{8}
 \nonumber\\
 &
 -344{r}^{6}{\Lambda}^{2}{r_h}^{3}-209 {r}^{5}{\Lambda}^{2}{r_h}^{4}
 +80 {r}^{7}{r_h}^{2}{\Lambda}^{2}-63 {{\Lambda}}^{2}{r_h}^{9}+132 \Lambda {r_h}^{7}
 +162 {r}^{3}{r_h}^{2}-12 {r_h}^{6}C
 \nonumber\\
 &
 +6 {r_h}^{2}{\Lambda} {r}^{5}-36 {r}^{4}C {r_h}^{2}-36 {r}^{2}C {r_h}^{4}
 +4 {r_h}^{8}C \Lambda-12 {r}^{6}C+4 {r}^{8}C \Lambda-132 {\Lambda} {r_h}^{4}{r}^{3} 
 \Big) r_b^6
 \nonumber\\
 &+
 \beta {r_h}^{2} \Big( 1488 {r}^{6}\Lambda r_h+16 {r}^{9}{\Lambda}^{2}
 -414 {r}^{4}r_h+16 {\Lambda}^{2}{r}^{8}r_h+27 {r_h}^{5}+312 {r}^{7}\Lambda
 -198 r{r_h}^{4}
 \nonumber\\
 &
 +558 {r}^{2}{r_h}^{3}-24 \Lambda {r_h}^{6}r +390 \Lambda {r_h}^{5}{r}^{2}
 +10 r C {\Lambda} {r_h}^{7}+54 {r}^{5}C {r_h}^{3}{\Lambda}+22 {r}^{7}C \Lambda r_h
 +42 {r}^{3}C \Lambda {r_h}^{5}
 \nonumber\\
 &+52 {r}^{2}C {\Lambda} {r_h}^{6}+96 {r}^{4}C {r_h}^{4}\Lambda
 +76 {r}^{6}C {r_h}^{2}\Lambda-9 {r}^{5}+1896 {\Lambda} {r_h}^{3}{r}^{4}
 -1310 {r}^{4}{\Lambda}^{2}{r_h}^{5}-448 {r}^{2}{\Lambda}^{2}{r_h}^{7}
 \nonumber\\
 &
 -268 {r}^{3}{{\Lambda}}^{2}{r_h}^{6}-78 r {\Lambda}^{2}{r_h}^{8}-
 1080 {r}^{6}{\Lambda}^{2}{r_h}^{3}-651 {r}^{5}{{\Lambda}}^{2}{r_h}^{4}
 -64 {r}^{7}{r_h}^{2}{\Lambda}^{2}-75 {\Lambda}^{2}{r_h}^{9}+144 \Lambda {r_h}^{7}
 \nonumber\\
 &
 +810{r}^{3}{r_h}^{2}-30 {r_h}^{6}C-132 {r_h}^{2}\Lambda {r}^{5}-162 {r}^{4}C {r_h}^{2}
 -126 {r}^{2}C {r_h}^{4}+10 {r_h}^{8}C {\Lambda}-66 {r}^{6}C+22 {r}^{8}C \Lambda
 \nonumber\\
 &
 -906 \Lambda {r_h}^{4}{r}^{3} 
 \Big) r_b^5
 \nonumber\\
 &
 -\beta {r_h}^{3} \Big( 1944 {r}^{6}\Lambda r_h+64 {r}^{9}{\Lambda}^{2}-324 {r}^{4}r_h
 +64 {\Lambda}^{2}{r}^{8}r_h+9 {r_h}^{5}+792 {r}^{7}\Lambda-162 r {r_h}^{4}+216 {r}^{2}{r_h}^{3}
 \nonumber\\
 &
 +6 \Lambda {r_h}^{6}r+354 \Lambda {r_h}^{5}{r}^{2}+8 r C {\Lambda} {r_h}^{7}
 +96 {r}^{5}C {r_h}^{3}{\Lambda}+50 {r}^{7}C \Lambda r_h
 +54 {r}^{3}C \Lambda {r_h}^{5}+62 {r}^{2}C {\Lambda} {r_h}^{6}
 \nonumber\\
 &
 +150 {r}^{4}C {r_h}^{4}\Lambda+146 {r}^{6}C {r_h}^{2}\Lambda-135 {r}^{5}
 +1800 \Lambda {r_h}^{3}{r}^{4}-1240 {r}^{4}{\Lambda}^{2}{r_h}^{5}
 -346 {r}^{2}{\Lambda}^{2}{r_h}^{7}-76 {r}^{3}{\Lambda}^{2}{r_h}^{6}
 \nonumber\\
 &
 -32 r {\Lambda}^{2}{r_h}^{8}-1416 {r}^{6}{\Lambda}^{2}{r_h}^{3}
 -819 {r}^{5}{{\Lambda}}^{2}{r_h}^{4}-432 {r}^{7}{r_h}^{2}{\Lambda}^{2}
 -33 {\Lambda}^{2}{r_h}^{9}+48 \Lambda {r_h}^{7}+1332 {r}^{3}{r_h}^{2}
 \nonumber
 \\
 &
 -24 {r_h}^{6}C-186 {r_h}^{2}\Lambda {r}^{5}-288 {r}^{4}C {r_h}^{2}
 -162 {r}^{2}C {r_h}^{4}+8 {r_h}^{8}C {\Lambda}-150 {r}^{6}C+50 {r}^{8}C \Lambda
 -1428 \Lambda {r_h}^{4}{r}^{3} \Big)
 r_b^4
 \nonumber\\
 &
 +\beta {r_h}^{4} \Big( 1212 {r}^{6}\Lambda r_h+100 {r}^{9}{\Lambda}^{2}+45 {r}^{4}r_h
 +100 {\Lambda}^{2}{r}^{8}r_h+1020 {r}^{7}\Lambda-63 r {r_h}^{4}
 -90 {r}^{2}{r_h}^{3}+6 \Lambda {r_h}^{6}r
 \nonumber\\
 &
 +174 {\Lambda} {r_h}^{5}{r}^{2}
 +2 r C \Lambda {r_h}^{7}+84 {r}^{5}C {r_h}^{3}\Lambda+60 {r}^{7}C \Lambda r_h
 +30 {r}^{3}C \Lambda {r_h}^{5}+32 {r}^{2}C \Lambda {r_h}^{6}+114 {r}^{4}C {r_h}^{4}\Lambda
 \nonumber\\
 & 
 +144 {r}^{6}C {r_h}^{2}\Lambda-225 {r}^{5}+510 \Lambda {r_h}^{3}{r}^{4}
 -551 {r}^{4}{\Lambda}^{2}{r_h}^{5}-136 {r}^{2}{\Lambda}^{2}{r_h}^{7}
 +68 {r}^{3}{\Lambda}^{2}{r_h}^{6}-3 r {\Lambda}^{2}{r_h}^{8}
 \nonumber\\
 &
 -936 {r}^{6}{{\Lambda}}^{2}{r_h}^{3}-495 {r}^{5}{\Lambda}^{2}{r_h}^{4}
 -616 {r}^{7}{r_h}^{2}{\Lambda}^{2}-6 {\Lambda}^{2}{r_h}^{9}+6 \Lambda {r_h}^{7}
 +1026 {r}^{3}{r_h}^{2}-6 {r_h}^{6}C-132 {r_h}^{2}\Lambda {r}^{5}
 \nonumber
\end{align}
\begin{align} 
 &
 -252 {r}^{4}C {r_h}^{2}-90 {r}^{2}C {r_h}^{4}+2 {r_h}^{8}C \Lambda-180 {r}^{6}C
 +60 {r}^{8}C \Lambda-1014 \Lambda {r_h}^{4}{r}^{3} \Big)
 r_b^3
 \nonumber\\
 &
 -\beta r {r_h}^{5} \Big( 40 {r}^{7}C {\Lambda}-120 {r}^{5}C+708 {r}^{6}\Lambda
 -9 {r_h}^{4}-90 {r_h}^{3}r+378 {r}^{2}{r_h}^{2}+76 {\Lambda}^{2}{r}^{8}-171 {r}^{4}
 -348 \Lambda {r_h}^{4}{r}^{2}
 \nonumber\\
 &
 +300 {r}^{5}r_h \Lambda+66 \Lambda {r_h}^{5}r
 -96 {r}^{4}\Lambda {r_h}^{2}-108 {r}^{3}C {r_h}^{2}-324 {r}^{5}{r_h}^{3}{\Lambda}^{2}
 -18 r C {r_h}^{4}
 -103 {r}^{3}{r_h}^{5}{\Lambda}^{2}
 \nonumber\\
 &
 -129 {r}^{4}{\Lambda}^{2}{r_h}^{4}
 -28 {\Lambda}^{2}{r_h}^{7}r+207 {r}^{3}r_h+{\Lambda}^{2}{r_h}^{8}
 -420 {r}^{6}{r_h}^{2}{\Lambda}^{2}-144 {r}^{3}\Lambda {r_h}^{3}
 +76 {r}^{7}{\Lambda}^{2}r_h
 \nonumber\\
 &
 +50 {r_h}^{6}{r}^{2}{{\Lambda}}^{2}+6 r C \Lambda {r_h}^{6}+42 {r}^{3}C \Lambda {r_h}^{4}
 +76 {r}^{5}C {\Lambda} {r_h}^{2}+40 {r}^{6}C \Lambda r_h+6 {r}^{2}C {r_h}^{5}\Lambda
 +36 {r}^{4}C \Lambda {r_h}^{3} \Big)r_b^2
 \nonumber\\
 &+
 \beta {r}^{2}{r_h}^{6} \Big( -144 {\Lambda}^{2}{r_h}^{2}{r}^{5}+28 {r}^{7}{\Lambda}^{2}
 -2 {r_h}^{7}{{\Lambda}}^{2}+6 {r}^{3}C {r_h}^{3}\Lambda-42 {r}^{4}C+14 {r}^{6}C \Lambda
 +28 {{\Lambda}}^{2}r_h {r}^{6}
 \nonumber\\
 &
 -54 \Lambda {r}^{3}{r_h}^{2}-48 {\Lambda} {r_h}^{4}r
 +14 {r}^{5}C \Lambda r_h-18 {r}^{2}C {r_h}^{2}+12 \Lambda {r_h}^{5}
 -12 \Lambda r_h {r}^{4}-63 {r}^{3}-18 {r_h}^{3}
 -96 \Lambda {r_h}^{3}{r}^{2}
 \nonumber\\
 &
 -56 {\Lambda}^{2}{r_h}^{3}{r}^{4}
 +20 {r}^{4}C \Lambda {r_h}^{2}+{\Lambda}^{2}{r_h}^{4}{r}^{3}+54 {r_h}^{2}r
 +10 {r_h}^{6}r {\Lambda}^{2}+108 {r}^{2}r_h+6 {r}^{2}C \Lambda {r_h}^{4}
 +252 {r}^{5}\Lambda \Big) r_b
 \nonumber\\
 &
 -\beta {r_h}^{7}{r}^{4} \Big( 36 \Lambda {r}^{3}-20 {\Lambda}^{2}{r_h}^{2}{r}^{3}
 +4 {\Lambda}^{2}r_h {r}^{4}+2 {r}^{3}C \Lambda r_h+4 {r}^{5}{{\Lambda}}^{2}
 +2 {r}^{4}C \Lambda+18 r_h-9 r+2 {r}^{2}C \Lambda {r_h}^{2}
 \nonumber\\
 &
 -12 {r_h}^{2}r\Lambda+5 {r_h}^{4}r {\Lambda}^{2}-12 \Lambda r_h {r}^{2}
 -4 {r_h}^{3}{r}^{2}{\Lambda}^{2}-12 {\Lambda} {r_h}^{3}+2 {r_h}^{5}{\Lambda}^{2}
 -6 {r}^{2}C \Big) ,
 \\
 &G_2(r)
 =12  \left( {\Lambda} {r_h}^{2}+{\Lambda} r_h r-3+{\Lambda} {r}^{2} \right)  
 \left( r_b-r_h \right) ^{4}
 \left( {r}^{2}r_b-{r}^{2}r_h+{r_h}^{2}r_b
 \right) ^{3},
\end{align}
where $C$ is the integration constant.

\section{Traceless condition}

The traceless condition gives a differential equation for $\xi(y)$. We obtain this differential equation in the original coordinate $r$:
\begin{align}
{\frac {d}{dr}}\xi \left( r \right) &=
{\frac { \left( \Lambda\,{r_h}^{3}-3\,r_h+6 r-4 \Lambda {r}^{3} \right) \xi
 \left( r \right) }{ \left( r-r_h \right) r  \left( {\Lambda} {r_h}^{2}+\Lambda r_h r-3+\Lambda {r}^{2}
 \right) }}
 +\frac{A}{B}h(r)
 +\frac{C}{D},
  \\
 A&=
 3 \big(  \left( \Lambda {r_h}^{3}+\Lambda r {r_h}^{2}+3 r_h+4 r_h \Lambda {r}^{2}
 +3 r \right) {r_b}^{2}
 - \left( 2 \Lambda {r_h}^{4} +6 {\Lambda} {r_h}^{2}{r}^{2} +6 r_h r +4 \Lambda {r}^{3}r_h \right) r_b
 \nonumber
 \\
 & \,\,\,\,\,\,\,\,
 -\Lambda {r_h}^{4}r+2 {\Lambda} {r_h}^{3}{r}^{2}
 +3 {r_h}^{2}r+2 \Lambda {r}^{3}{r_h}^{2} \big) r,
 \\
 B&=
 \left( r-r_h \right)  
 \left(  \left( {r}^{2}+{r_h}^{2} \right) r_b-{r}^{2}r_h \right)  \left( r_b-r_h \right)  
 \left( \Lambda {r_h}^{2}+\Lambda r_h r-3+\Lambda {r}^{2}\right) ^{2},
 \\
 C&= 
 -3r  \Big(  \left( -3 {r_h}^{2}r+11 \Lambda {r_h}^{4}r+11 \Lambda {r_h}^{5}-3 {r}^{3}
 +11 {\Lambda} {r_h}^{3}{r}^{2}-39 {r_h}^{3}-15 {r}^{2}r_h
 +3 \Lambda {r}^{3}{r_h}^{2} \right) {r_b}^{3}
 \nonumber\\
 &\,\,\,\,\,\,\,\,
 + \big( -4 \Lambda {r_h}^{6}-22 \Lambda {r_h}^{5}r
 +21 {r}^{2}{r_h}^{2}-4 r_h \Lambda {r}^{5}+54 {r_h}^{3}r-4 \Lambda {r_h}^{2}{r}^{4}
 -25 \Lambda {r}^{3}{r_h}^{3}+33 {r}^{3}r_h
 \nonumber\\
 &\,\,\,\,\,\,\,\,
 -19 \Lambda {r_h}^{4}{r}^{2}+18 {r_h}^{4} \big) {r_b}^{2}
 + \big( -45 {r}^{3}{r_h}^{2}+21 \Lambda {r}^{3}{r_h}^{4}
 -6 {r_h}^{3}{r}^{2}-27 {r_h}^{4}r +9 {r_h}^{6}\Lambda r
 +6 {\Lambda} {r_h}^{5}{r}^{2}
 \nonumber\\
 &\,\,\,\,\,\,\,\, 
 +6 \Lambda {r_h}^{3}{r}^{4}+6 \Lambda {r}^{5}{r_h}^{2} \big) r_b
 -2 {\Lambda} {r}^{5}{r_h}^{3}+15 {r}^{3}{r_h}^{3}-2 {\Lambda} {r_h}^{4}{r}^{4}
 -5 \Lambda {r_h}^{5}{r}^{3} \Big),
 \\
 D&=
 -\left( r-r_h \right)\left( r_b-r_h \right)  \left( \Lambda {r_h}^{2}
 +\Lambda r_h r-3+\Lambda {r}^{2} \right) ^{2} 
 \left(  \left( {r}^{2}+{r_h}^{2} \right) r_b-{r}^{2}r_h \right) ^{2}.
\end{align}
We solve this equation near the horizon ($r=r_h$) and the boundary ($r=r_b$) and obtain the asymptotic 
solutions (\ref{asym2}) and (\ref{asym3}) using the relation (\ref{transe}).

\bigskip\bigskip

\end{document}